\input{psfig.sty}
\documentclass[useAMS]{mn2e}
\usepackage{graphicx}
\usepackage{subfigure}
\title[Surface Density of dark matter haloes on galactic and cluster scales]{Surface Density of dark matter haloes on galactic and cluster scales}

\author[A. Del Popolo, V.F. Cardone, G. Belvedere]{A. Del Popolo$^{1,2}$\thanks{Corresponding author\,: {\tt adelpopolo@oact.inaf.it}}, V.F. Cardone$^3$, 
G. Belvedere$^1$ \\
$^1$Dipartimento di Fisica e Astronomia, Universit\`{a} di Catania, Viale Andrea Doria 6, 95125 Catania, Italy \\
$^2$ Departamento de Astronomia, Universidade de Sao Paulo, Rua do Matao 1226, 05508-900 Sao Paulo, SP, Brazil\\
$^3$I.N.A.F. - Osservatorio Astronomico di Roma, via Frascati 33, 00040 - Monte Porzio Catone (Roma), Italy 
}

\begin{document}

\pagerange{\pageref{firstpage}--\pageref{lastpage}} \pubyear{2002}

\maketitle

\label{firstpage}

\begin{abstract}

In this paper, in the framework of the secondary infall model, the correlation between the central surface density and the halo core radius of galaxy, and cluster of galaxies, dark matter haloes was analyzed, this having recently been studied on a wide range of scales. 
We used Del Popolo (2009) secondary infall model taking into account ordered and random angular momentum, dynamical friction, and dark matter (DM) adiabatic contraction to calculate the density profile of haloes, and then these profiles are used to determine the surface density of DM haloes.
The main result is that $r_\ast$ (the halo characteristic radius) is not an universal quantity as claimed by Donato et al. (2009) and Gentile et al. (2009). On the contrary, we find a correlation with the halo mass $M_{200}$ in agreement with Cardone \& Tortora (2010), Boyarsky at al. (2009) and Napolitano et al. (2010), but with a significantly smaller scatter, 
namely $0.16 \pm 0.05$. 
We also consider the baryon column density finding this latter being indeed a constant for low mass systems such as dwarfs, but correlating with mass with a slope $\alpha= 
0.18 \pm 0.05$. In the case of the surface density of dark matter for a system composed only of dark matter, as in dissipationless simulations, we get $\alpha=0.20 \pm 0.05$. These results leave little room for the recently claimed universality of (dark and stellar) column density.
\end{abstract}


\section{Introduction}

The $\Lambda$CDM model is remarkably successful in fitting a wide range of data on cosmological scales, from the Hubble diagram of Type Ia Supernovae (Kowalsky et al. 2008) to the matter power spectrum with its Baryonic Acoustic Oscillation features (Percival et al. 2010) and the anisotropy and polarization spectrum of cosmic microwave background radiation (Komatsu et al. 2011). Notwithstanding this long list of successes awarding the $\Lambda$CDM scenario the name of concordance model, there is still some tension at scales from a few kpc to tens of pc so that great amount work has been recently dedicated to test what $\Lambda$CDM predicts for galaxies and clusters. On the one hand, numerical CDM simulations typically form a number of satellite haloes, i.e. small haloes and galaxies orbiting around a massive central system, which is far larger than what is indeed observed. As a clear evidence of this {\it substructure problem} (Moo et al. 1999), one can note that the best studied satellite systems of the Milky Way and Andromeda galaxies are hard to reconcile with the predicted abundance of small\,-\,mass dark matter clumps even if one takes into account detection effects. On the other hand, rotation curves of spiral galaxies favor cored dark halo models in contrast with the cusps predicted by the $\Lambda$CDM model de Blok et al. (2009), 
while a severe discrepancy is also observed for what concerns the size and the angular momenta of galactic disks (Mayer et al. 2008). Although both theoretical models and increasingly accurate observations have been put down during the past 10\,-\,15 years, the quoted problems still stay on the ground. Finding the origin of such discrepancies is complicated mainly due to the lack of a well understood method to describe the complex hydrodynamical phenomena in hot dense plasma and physical interactions of baryons and radiation which follow the initial (well understood) growth and collapse of the initial density perturbations. 
As a valuable help to address this problem, one can look for scaling relations
among DM haloes parameters and stellar quantities in order to better constrain the formation scenarios and the DM
properties. 
In this context, an intriguing property of dark matter haloes was noted by Kormendy \& Freeman (2004) based on halo parameters obtained by mass modeling 55 spiral galaxy rotation curves within the framework of the Maximum Disk Hypothesis (MDH).
%
%
Among other relations between the halo parameters, they found that the quantity $\mu_{0D} =\rho_0 r_0$\footnote{$\rho_0$ and $r_0$
are, respectively, the central density and core radius of the adopted pseudo-isothermal cored dark matter density profile. 
}, proportional to the halo central surface density for any cored halo distributions, is nearly independent of the galaxy blue magnitude.  In particular, they found that this
quantity takes a value of $\simeq 100 M_{\odot}/pc^{-2}$. Donato et al. (2009, D09), 
by means of rotation curves of $\simeq  1000$ spiral galaxies, the mass models of individual dwarf and spiral galaxies and the weak lensing signal of elliptical and spirals,
found strong evidences for the constancy of the central DM column density, over 12 orders of magnitude in luminosity. They found $\log \mu_{0D}=2.15 \pm 0.2$, in units of $\log (M_{\odot}/pc^2)$.
Gentile et al. (2009, G09) extended the result to the luminous matter surface density, finding that total luminous--to--dark matter ratio
within one halo scale-length is constant. 
Opposite results were obtained by Napolitano, Romanowsky \& Tortora (2010, NRT10) that showed that, on average, the projected density of local
early type galaxies (ETGs), within effective radius is systematically higher than the same quantity for spiral and dwarf galaxies, pointing
to a systematic increase with halo mass as suggested by Boyarsky et al. (2009a, B09). These last extended the samples
above analyzed to both group and cluster scale systems finding that the {\em dark matter column density}, $S$, (defined in Sect. 3 and equivalent to $\mu_{0D}$ in the case of fit with Burkert profile (Burkert 1995)) is given by
\begin{equation}
\log S= 0.21 \log \frac{M_{halo}}{10^{10} M_{\odot}}+1.79
\end{equation}
with S in $M_{\odot}/pc^2$. 
Cardone \& Tortora (2010) (CT10), by modeling the dark halo with a Navarro - Frenk -White profile and assuming a Salpeter initial mass
function (IMF) to estimate stellar masses, found that the column density and the Newtonian acceleration within the halo characteristic radius $r_s$ and effective radius
$R_{eff}$ are not universal quantities, but correlates with the luminosity $L_V$ , the stellar mass $M_{\ast}$ and the halo mass $M_{200}$, contrarily to what is expected from D09 and G09, and in agreement with B09.

In order to try to discriminate among these results and find an explanation and analytical derivation of the surface density of halos, we analyze the problem using
the secondary infall model (SIM) introduced in Del Popolo (2009) (hereafter DP09), taking into account ordered and random angular momentum, dynamical friction, and baryon adiabatic contraction. 
The plan of the paper is the following. In Sect. 2, we summarize the model. In sect. 3, we discuss the results and Sect. 4 is devoted to conclusions.

\section{Model}

The above discussed results about a constant surface density (D09; G09) of DM together with the ones suggesting a mass dependence (B09; CT10) are substantially grounded on fitting observational properties. 
With the exception of Boyarsky et al. (2010) (B10), no qualitative explanation and/or analytical derivation of the quoted results  
has been proposed so far. B10 used a SIM and supposed that the halo density profile evolves in a self--similar way as previously done by Fillmore \& Goldreich (1984) (FG84), Bertschinger (1985) (B85) and Skivie et al. (1997) (S97). 
They arrived at the conclusion that $S \propto M^{1/3} t^{-4/3}$ qualitatively in agreement with B09 simulations and CT10 results, in the sense that $S$ increases with mass, but the slope is much bigger than what obtained by B09 and CT10. 

The quoted discrepancy could be connected to the oversimplified structure of the SIM used. Even if they tried to use the improved SIM of S97, taking account angular momentum,
these effects on collapse, in reality, are not properly taken into account. 

In the following, we will shortly describe how implementing the SIM approach in oirder to obtain the dark matter halo density profile. The interested reader can find more in DP09.

The halo profiles are derived by using the improved secondary infall model of DP09. In the model, a bound mass shell of initial comoving radius $x_i$ expands up to a maximum radius (or turnaround radius) $x_{\rm ta}$. As successive shells expand, they acquire angular momentum and then contract on orbits determined by the angular momentum itself, while dissipative processes and eventual violent relaxation intervene to virialize the system
converting kinetic energy into random motions. The final density profile may then be computed as\,:

\begin{equation}
\rho(x) = \frac{\rho_{\rm ta}(x_{\rm ta})}{(x/x_{\rm ta})^3} \left [ 1 + \frac{d\ln{(x/x_{\rm ta})}}{d\ln{x_{\rm ta}}} \right ]
\label{eq: defrhoend}
\end{equation}
with $\rho_{\rm ta}(x_{\rm ta})$ the density at turnaround and $x/x_{\rm ta}$ referred to as the collapse factor (see Eq. A18 in DP09). To describe the proto--haloes density profile, DP09 considered the profile of a peak in the density field generated according to the matter power spectrum of Bardeen et al. (1986) and then took into account angular momentum, dynamical friction and the presence of baryons as described below.

First, the angular momentum is decomposed in an ordered component, related to the tidal torques experienced by proto--haloes, and a random component connected to random velocities (Ryden \& Gunn 1987). The ordered term is computed following Ryden (1988a,b), while the random part is assigned to proto\,-\,structures according to Avila\,-\,Reese et al. (1998) and Ascasibar et al. (2004). A term related to the dynamical friction force has been explicitly introduced in the equations of motion and evaluated Kandrup (1980) dividing the gravitational force into an average and a random component generated by the clumps in the hierarchical universe. Finally, adiabatic contraction of the halo, due to the baryonic collapse, has been taken into account through the formalism of Klypin et al. (2002) and Gnedin et al. (2004), also including the exchange of angular momentum among baryons and dark matter (Klypin et al. 2002, and Gnedin et al. 2004).

As previously reported, the details of the model are given in DP09. The profile formation is due to a complex  
baryon-dark matter (DM) interplay that can be summarized as follows

 Initially the proto-structure is in the linear phase, it expands, reach a maximum
of expansion and then collapse. Baryons trapped inside the potential wells of DM halos are subject
to radiative dissipation processes which give rise to clumps and self-gravitating clouds before it collapses
to the halo center and condenses into stars and galaxies. The stage of baryons cooling and stars
formation happens as described in Ryden (1988) (Sect. 4: "baryionic dissipation"). In the infall, the
baryons compress the DM halos (AC), producing a steepening of the DM density profile. When clumps
reaches the central high density regions they experience a dynamical friction force from the less massive
DM particles as they move through the halo. Dynamical friction acts as an angular momentum
engine (Tonini, Lapi \& Salucci 2006), and energy and angular momentum are transferred to 
DM (El-Zant et al. 2001), increasing its random motion, and giving rise to a motion of DM particles
outwards. Moreover, ordered angular momentum mainly acquired in the expansion phase, through
tidal torques, gives rise to nonradial motions in the collapse phase that amplifies the effects of the previous
mechanism. Then the joint effect of angular momenta (ordered and random) and dynamical friction trasporting it from baryons to DM, overcomes
that of the AC and the profile starts to flatten.

The comparison of the quoted model with SPH simulations has been performed in several papers (e.g., Del Popolo 2012, Fig. 3), showing 
a good agreement with hydrodinamical simulations.


The final product of this halo formation method gives the DM density profile as function of the radius $r$ and the total halo mass $M_{\rm vir}$. The latter is the only parameter needed in order to specify the halo density, being the halo inner slope a function of the virial mass as well.
Such a dependence breaks the universality of the halo profiles and favor Burkert models at dwarf scales, and models steeper than NFW at normal galaxy scales.

\section{Results}

\subsection{The S-M relation}

D09 and G09 analyzed rotation curves and weak lensing data for a sample of dwarf, spiral and elliptical galaxies fitted by the Burkert profile:
\begin{equation}
\rho(r)_B=\frac{\rho_0r_0^3}{(r+r_0)(r^2+r_0^2)}
\label{eq:bur}
\end{equation}
The fit to the rotations curves yields the values of the two structural DM parameter (i.e., $r_0$ and $\rho_0$), then the surface density is calculated as $\mu_{0D}=\rho_0 r_0$. 
The previous approach is based on the assumption that the galaxies studied, namely dwarf, spiral and elliptical galaxies can be fitted by cored models
\footnote{The existence of a constant central surface density of DM in galaxies is independent from the core model used (D09). 
The choice of of a Burkert profile (which is a model with core) is due to the fact that at radii $>0.3 r_{vir}$ (where $r_{vir}$ is the virial radius) the Burkert profile and the NFW profile converge, while at small radii and for appropriate values of $r_0$ it reproduce the NFW profile. }.
Now, the D09 and G09 samples contains dwarfs, spirals and ellipticals. 

For what concerns dwarfs, a large part of the studies of their density profiles indicate that 
the inner part of density profiles is characterized by a core-like
structure (e.g., Flores \& Primak 1994; Moore 1994;
Kravtsov et al. 1998; de Blok \& Bosma 2002; de Blok, Bosma \& McGaugh 2003; Gentile et al.
2004, 2007; Blaise-Ouelette et al. 2004, Span\'o et al. 2008,
Kuzio de Naray et al. 2008, 2009, and Oh et al. 2010), 
but some studies found that density profiles
are cuspy. 
Other studies concluded that density profiles are compatible with both cuspy and
cored profiles (van den Bosch et al. 2000; 
Simon et al. 2005; de
Blok et al. 2008). Very interesting are the results of Simon et al. (2005), and de Blok et al. (2008).
The first showed that the power-law index, $\alpha$, describing the central density profile, span the range from
$\alpha = 0$ (NGC 2976) to $\alpha= 1.28$ (NGC 5963), with a mean
value $\alpha = 0.73$. Del Popolo (2012), showed how cuspy profiles can form even in the case of dwarfs.
The second (de Blok et al. 2008), by using the THINGS sample,
found that 
for low mass galaxies a core dominated halo is clearly preferred over a cusp-like halo, while for massive, disk dominated galaxies, the two models
fit apparently equally well. In other terms, while a large part of dwarfs are well fitted by core-like profiles, some of them are not (see also Del Popolo 2012). {If this cannot be taken as an evidence in favour of cusps, it shows that using core profile for every spiral is not always correct, and if one is using a sample containing also ellipticals the problem is even more serious. Moreover, THINGS shows that there is a trend to more cuspy profiles going to higher masses. The trend is confirmed by the fact that ellipticals are characterized by cuspy profiles and similarly clusters of galaxies
have mainly cuspy profiles. (see the following of the discussion). \\}
NRT10 studied the DM surface density finding that their result is in good agreement with those of D09, namely they found a constant surface density on average in the mass range going from dwarfs to late types, but differently from D09 they found a  
systematic increase with mass as found by Boyarsky et al. (2009). Moreover, according to them, the consistency of the NFW models with the dwarf
and spiral data shown in their Fig. 9 suggests that the surface density result 
does not necessarily imply cored DM haloes.

In the case of ellipticals, the situations is even more complicated. Several studies (e.g., Mamon \& Lokas 2005) showed that the DM profile of ellipticals is cuspy. 
NRT10 found that early-type-galaxie (ETG) violate
the constant density scenario for the other galaxies by
a factor of $\sim 10$ on average, and a factor of $\sim  5$ in the same
mass regime, in agreement with Gerhard et al. (2001),
Thomas et al. (2009), and  Boyarsky et al. (2009).

The quoted discussion imply that when we study the surface density it is preferable to have a ``method" that is able to fit not only 
cored density profiles but also cuspy or intermediate among the two types. 


Alternatively we can determine the surface density, in a more general way than that used by D09 and G09, by means of a {dark \emph matter column density}, which represents an average over the central part of the object: 
\begin{equation}    
\label{eq:Sbar}
S = \frac2{r_\star^2} \int^{r_\star}_0 rdr
\int  dz \rho_{DM}(\sqrt{r^2+z^2})
\label{eq:column}
\end{equation}

The integral over $z$ is extended to the vitial of the halo. Definition ~(\ref{eq:Sbar}) has as a consequence that $S$ scales as the dark
matter surface density within $r_\star$ ($ S \propto \rho_\star r_\star$), 
\footnote{Parameters of different profiles that fit the same DM density distribution are related (for example, $r_s$ for NFW is equal to $6.1r_c$ for ISO and equals to $1.6r_B$ for BURK). Choosing these values as $r_\star$ in each case, one finds that the value of $S$ for NFW 
and ISO 
differ by less than 10\% (the difference
between NFW and BURK is $\sim 2\%$), see Supporting Information for details.}
but $S$ is more universal, being defined for any density profile (not necessarily in the core). 
In other terms, if the profile is cored using S we will obtain the same result for the surface density of the Burkert's fit, while in the case the profile is not cored S will give a more precise value for the surface density than using the Burkert's model fit.

%
%

The quoted method was already used by B09 who extended the analysis of D09 and G09 to galaxies and galaxy clusters and fitted the DM profiles by means of three DM profiles models, namely Burkert profile (Eq. \ref{eq:bur}), pseudo-isothermal (ISO) profile

\begin{equation}
\rho(r)_{ISO}=\frac{\rho_c}{1+r^2/r_c^2}
\end{equation}
where $r_c$ is the ISO core radius,
and NFW profile:
\begin{equation}
\rho(r)_{NFW}=\frac{\rho_s r_s}{r(1+r/r_s)^2}
\end{equation}
where $r_s$ is the typical NFW characteristic radius.

CT10 used two DM profiles for the fit, namely Burkert profile and ISO.

In order to determine $S$, we use Eq. (\ref{eq:column}) together with the density profiles obtained with the model of the previous section, which are plotted in
Fig. 1, 3, and 5 of DP09. As there shown, density profiles become monotonically shallower inwards, down to the innermost resolved point, with no indication
that they approach power-law behavior. This is in agreement with recent N--body simulations (e.g., Navarro et al. 2010).
Navarro et al. (2010) showed that density profiles deviate slightly but systematically from
the NFW model, and are approximated more accurately by a fitting formula where the logarithmic slope is a power-law of radius, the Einasto profile 
\begin{equation}
\ln(\rho(r)/\rho_{-2})=-2/\alpha [(r/r_{-2})^{\alpha}-1],
\end{equation}
where $r_{-2}$ and $\rho_{-2}$ are connected to the scaling radius and density of the NFW profile (see the following) by
$r_{-2}=r_s$ and $\rho_{-2}=\rho_s/4$.
A comparison between the results of our model and dissipationless N-body simulations and SPH simulations are shown in Del Popolo (2011).

\begin{figure}
\includegraphics[width=90mm]{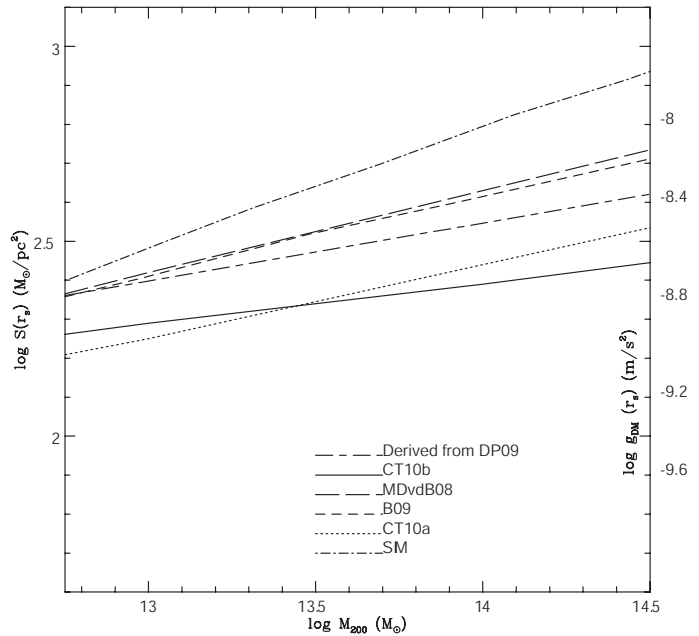}
\caption{${\cal{S}}(r_s)$ as a function halo mass $M_{200}$. From bottom to top we have that: the dotted line represents the best fit linear relation using
the D' Agostini (2005) fit, while the solid line the direct fit methods, obtained in CT10. 
The long--short dashed line represents ${\cal{S}}(r_s)$ obtained with the model described in the present paper. The short--dashed line represents prediction and the results from the $\Lambda$CDM
N--body simulation of Maccio et al. (2008). The long--dashed line represents the B09 best fit linear relation.
The dot--dashed line represents the secondary--infall model (B10) prediction.  
Black points and open boxes plot ${\cal{S}}(r_s)$ as a function halo mass $M_{200}$, respectively for the NFW+Salpeter model and when a Chabrier IMF is used, obtained by CT10. 
}
\end{figure}

\begin{figure}
\includegraphics[width=90mm]{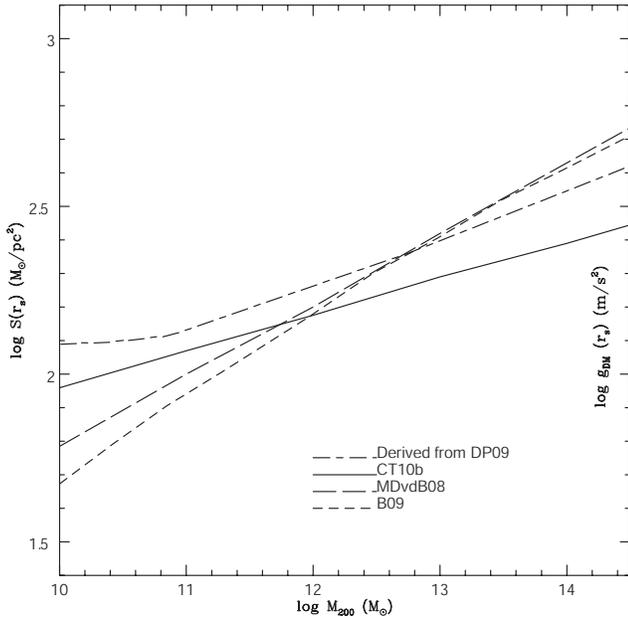}
\caption{${\cal{S}}(r_s)$ as a function halo mass $M_{200}$. From bottom to top we have that: the short--dashed line represents the prediction and the results from the $\Lambda$CDM N--body simulation of Maccio et al. (2008); the long--dashed line the B09 best fit linear relation; 
the solid line represents the direct fit methods, obtained in CT10. 
The short--long dashed line represents ${\cal{S}}(r_s)$ obtained with the model described in the present paper. 
}
\end{figure}

In Fig. 1, ${\cal{S}}(r_s)$ vs $M_{200}$ is plotted for our model and for different authors'.
From bottom to top: the best fit linear relation, found using the D' Agostini (2005) fit, is represented by the dotted line, while the solid line is related to the direct fit methods obtained in CT10. 
The long--short dashed line represents ${\cal{S}}(r_s)$ found in the framework of the model of this paper. The short--dashed line is connected to the prediction and results of the $\Lambda$CDM
N--body simulation by Maccio et al. (2008). The long--dashed line is related to the B09 best fit linear relation.
The dot--dashed line is related to the secondary--infall model (B10) theoretical prediction. 

The maximum likelihood fit for the correlation $\log \cal{S}$$(r_s)$-$\log M_{200}$ in this model is 
\begin{equation}
\log {\cal{S}}(r_s)=0.16 \log(\frac{M_{200}}{10^{12} M_{\odot}})+2.23
\end{equation}
while in the case of CT10
is
\begin{equation}
\log {\cal{S}}(r_s)=0.16 \log(\frac{M_{200}}{10^{12} M_{\odot}})+2.11
\end{equation}
and a steeper slope is obtained with the stellar mass, $M_{\ast}$
\begin{equation}
\log {\cal{S}}(r_s)=0.29 \log(\frac{M_{200}}{10^{11} M_{\odot}})+2.17
\end{equation}
The marginal constraints on the scaling relationships parameters for the correlation  
$\log {\cal{S}}(r_s)$-$\log M_{200}$ is $0.16^{+0.05}_{-0.05}$ in this model,
while for CT10, considering a fiducial NFW+Salpeter model, 
is $0.14^{+0.15}_{-0.15}$ and for the correlation $\log {\cal{S}}(r_s)-\log M_{\ast}$ is $0.29^{+0.15}_{-0.15}$ (Table 3 of CT10).

%
%

\begin{figure}
\vskip 15.5cm 
\hspace{-3.5cm} 
{\includegraphics{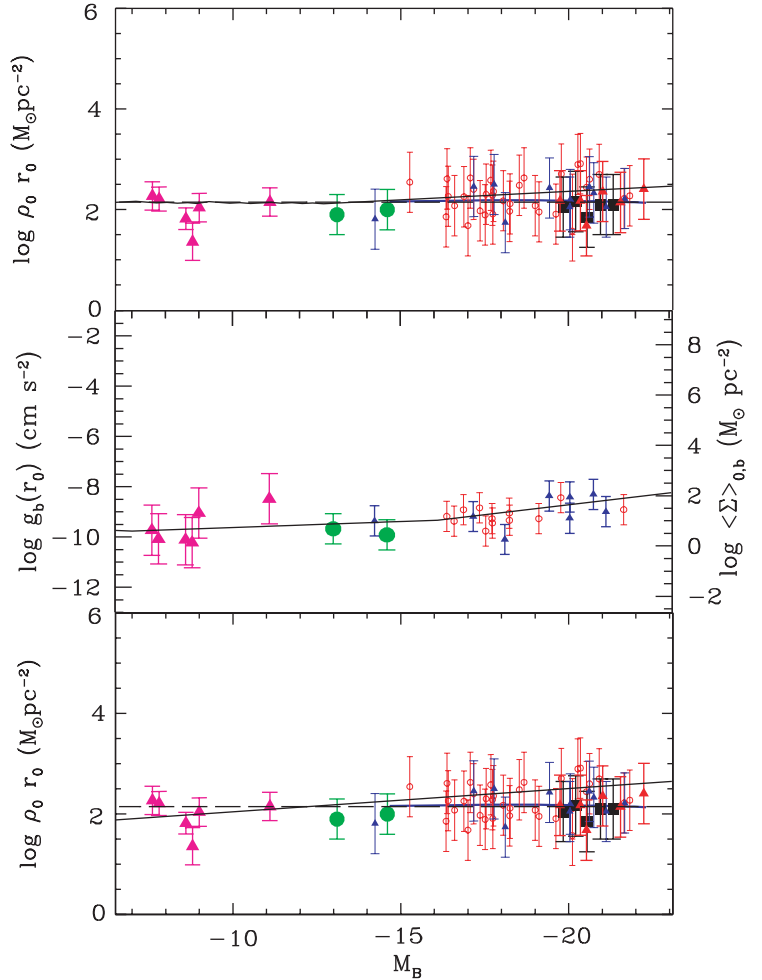}}
\vskip -2.5cm 
\caption[]{{\bf a} \small $\rho_0 r_0$ in units of $M_{\odot}/pc^2$ as a function of galaxy magnitude for different galaxies and Hubble Types. Data are from: Spano et al. (2008) data (empty small red circles), 
the dwarf irregulars (full green circles) N 3741 ($M_B = -13.1$) and DDO 47 ($M_B = -14.6$), Spirals and Ellipticals investigated by weak lensing (black squares), dSphs (pink triangles), nearby spirals in THINGS (small blue triangles), and early-type spirals (full red triangles). The long dashed line is the result of D09. The solid line is the result of the present paper. 
{\bf b} $<\Sigma>_{0,bary}$ and $g_{bary}(r_0)$ are plotted as a function of the B-band absolute magnitude of the galaxies. From the original sample of D09, here, as in G09, are used the dwarf spheroidals data, NGC 3741, DDO 47, and the two samples of spiral galaxies (Spano et al. 2008; de Blok et al. 2008) which all together span the whole magnitude range probed in the original sample (D09). The magenta triangles are the dwarf spheroidal galaxies1, the left green point is NGC 3741 and the right green point is
DDO 47, and the empty red circles and filled blue triangles symbols are the Spano and THINGS  (de Blok et al. 2008) spiral galaxies samples, respectively.
{\bf c} Same as panel a, but in this case our $\rho_0 r_0$ is calculated without taking account of baryons in our SIM, as in dissipationless simulations.
}
\end{figure}

The best fit concerning DM halos in CT10 is shallower than the B09 and the result of the present paper, altough the slope is consistent with B09 (0.21) and the present paper within the large error bars. One has to recall that while CT10 have explicitly taken into account the
correlated errors on $S(r_s)$ and $M_{200}$, it is not known if 
B09 does the same. 
Then, the difference in slope is only the outcome of the
use of different algorithms on noisy data. 
However, if we compare the results of the present paper, those of B09 (over
the same mass range) the estimates from the $\Lambda$CDM N- body simulations
of isolated haloes from Macci\'o, Dutton \& van den Bosch (2008) and the predictions from the secondary infall model (B09), are on average larger than those of CT10 $S(rs)$ values. 
A possible reason for the difference 
that the larger values in literature 
neglecting the stellar component. 

Another fundamental point to remark is the fact that in CT10, because of the large errors the 
$\log {\cal{S}}$$(r_s)$-$\log M_{200}$ is consistent even with a slope $0$. In our paper, because of the smaller errors the flat behavior of the $S-M$ relation is not allowed. This result is of great importance for two reasons that we will describe in the following (see section 3.2).
 
In their paper, CT10 conclude that results for their reference model argue against the universality of the column density proposed in D09. However, the same authors showed that describing the DM halo adopting a Burkert model and a Chabrier IMF (similar to those used in D09), one obtains that 
\begin{equation}
\log {\cal{S}}(1.6 r_B)=0.02 \log(\frac{L_{V}}{10^{11} L_{\odot}})+2.65
\end{equation}
implying that $S(1.6 r_B)$ is constant with $L_V$
but using a Salpeter IMF they got $S \propto L_V^{0.20}$, arguying against universality and for a strong model dependency. 

In Fig. 2, we plot $S$ versus $M_{200}$ for masses in the range $10^{10}$-$10^{14.5} M_{\odot}$. The plot uses the same symbols of Fig. 1, but now for clarity we did not plot the SIM model of B10, the best fit linear relation using
the D' Agostini (2005) fit. The plot shows that CT10 fit gives larger values than the Macci\'o et al. (2008) and B09 N-body simulations. The $S-M$ relation obtained from the present paper shows a flattening for masses smaller than $5 \times 10^{10} M_{\odot}$. This flattening is the result of the fact that for dwarf--galaxies the density profiles tend to become shallower and fit with Burkert's profile are better than those obtained through NFW model. 
This agrees with the claim of D09 and G09. However, we want to stress that in the case of D09 and G09, because of the large error bars the claim of constancy of the surface density for all masses is not univocal. It is very easy to show that the points in D09 and G09 can be easily fitted by a line with slope of $0.16$ ($S \propto M^{0.16}$). 
Comparing the results with N-body simulations (B09 and Macci\'o et al. 2009) shows that for larger masses in the cluster mass region, the prediction for $S$ of the present paper are in agreement with simulations but for smaller masses tend to diverge. The reason of this is due to the fact that N-body simulations uses just CDM particles and do not take account of the effects of baryons and the others taken into account in the present paper model.

In order to compare our results to those of D09, and G09, we need to find a way to express our result in terms of $\mu_{0D}=\rho_0 r_0$. 

In the following, we show some examples for known profiles. Before we write the $S(r)$ for them. 
For the  pseudo-isothermal profile one obtains:
\begin{equation}
  S_{ISO}(R) =\frac{2\pi\rho_c r_c^2}{R^2}\left[\sqrt{R^2+r_c^2}-r_c\right].
  \label{eq:9}
\end{equation}
For the NFW density distribution, we have
\begin{equation}
S_{NFW}(R) = \frac{4\rho_s
    r_s^3}{R^2}\left[\frac{\arctan\sqrt{R^2/r_s^2-1}}{\sqrt{R^2/r_s^2-1}} +
    \log\left(\frac{R}{2r_s}\right)\right]\label{eq:S_NFW}\;.
\end{equation}
Analytical relation between parameters of several profiles can be obtained fitting the same rotation curve (B09, CT10).
For an ISO and NFW profile, one can take an ISO profile, calculates the relative rotation curve and fits the result 
with a NFW profile. One find: $r_s \simeq 6.1 r_c$, $\rho_s \simeq 0.11 r_c$.
Comparing the column densities for NFW and ISO profiles, whose parameters are related through the relations now written, one obtains: 
\begin{equation}
\label{eq:15}
\frac{S_{NFW}(r_s)}{S_{ISO}(6r_c)} \approx 0.91 \quad .
\end{equation}

Similarly for the Burkert profile and NFW: $r_s \simeq 1.6 r_0$, $\rho_s \simeq 0.37 \rho_0$ and
\begin{equation}
\label{eq:15}
\frac{S_{NFW}(r_s)}{S_{Burkert}(1.66r_0)} \approx 0.98 \quad .
\end{equation}

The difference between the column densities $S_{NFW}$, $S_{Burkert}$ and $S_{ISO}$ turns out to be less than $10\%$ (CT10). 

Then $S_{NFW}(r_s) \simeq 0.98 S_{Burkert}(1.6r_0) \approx 1.89 r_0 \rho_0$.

In Fig. 3a, Fig. 3b, and Fig. 3c, we compare the result of the present model with D09 and G09 results. 
Fig. 3a plots $\rho_0 r_0$ in units of $M_{\odot}/pc^2$ as a function of galaxy magnitude for different galaxies and Hubble Types\footnote{The empty small red circles are taken from Spano et. al. (2008).
The URC (solid blue line), the dwarf irregulars (full green circles) N 3741 ($M_B = -13.1$) and DDO 47 ($M_B = -14.6$), Spirals and Ellipticals investigated by weak lensing (black squares), dSphs (pink triangles), nearby spirals in THINGS (small blue triangles), and early-type spirals (full red triangles). The long dashed line is the result of D09. The solid line is the result of the present paper. }.
The quoted figure shows that the column density is indeed constant for masses smaller than $\sim 5 \times 10^{10} \ {\rm M_{\odot}}$ and luminosities smaller than $M_B \sim -14$. However, a systematic change of the average column density $\rho_0 r_0$ as a function of the object mass is clearly present for larger masses, the data being well fitted as $\rho_0 r_0 \propto M^{\alpha}$ with $\alpha=0.16 \pm 0.05$. Such results make, therefore, us safely argue against the universality of ${\cal{S}}_{DM}$ (i.e., ${\cal{S}}_{DM} = const$) claimed by D09.

Fig. 3b shows instead the stellar column density denoted here $\langle \Sigma \rangle_{0b}$ to be consistent with G09\footnote{From the original sample of D09, here, as in G09, are used the dwarf spheroidals data, NGC 3741, DDO 47, and the two samples of spiral galaxies (Spano et al. 2008; de Blok et al. 2008) which all together span the whole magnitude range probed in the original sample (D09). The magenta triangles are the dwarf spheroidal galaxies1, the left green point is NGC 3741 and the right green point is
DDO 47, and the empty red circles and filled blue triangles symbols are the Spano et al.(2008) and THINGS (de Blok et al 2008) spiral galaxies samples, respectively.}.
For this same reason, we also show the corresponding Newtonian acceleration, $g_{b}(r_0) = \pi G \langle \Sigma \rangle_{0b}$ in $cm/s^2$. Both quantities are plotted as a function of the $B$ band absolute magnitude using the same dataset in G09 which is obtained from the D09 sample removing some problematic objects (see G09 for details). As for ${\cal{S}}_{DM}$, we again find that $\langle \Sigma \rangle_{0b}$ is constant for $M \le 5 \times 10^{10} \  {\rm M_{\odot}}$, while it scales with $M$ as $M^{\alpha}$ with $\alpha = 0.18 \pm 0.05$ for larger masses. We therefore agree with both D09 and G09 at low masses, but the slope at high $M$ is a clear evidence against their claim that the baryons acceleration at $r_0$ is constant over the full mass range.

\begin{figure*}
\includegraphics[width=12.25cm]{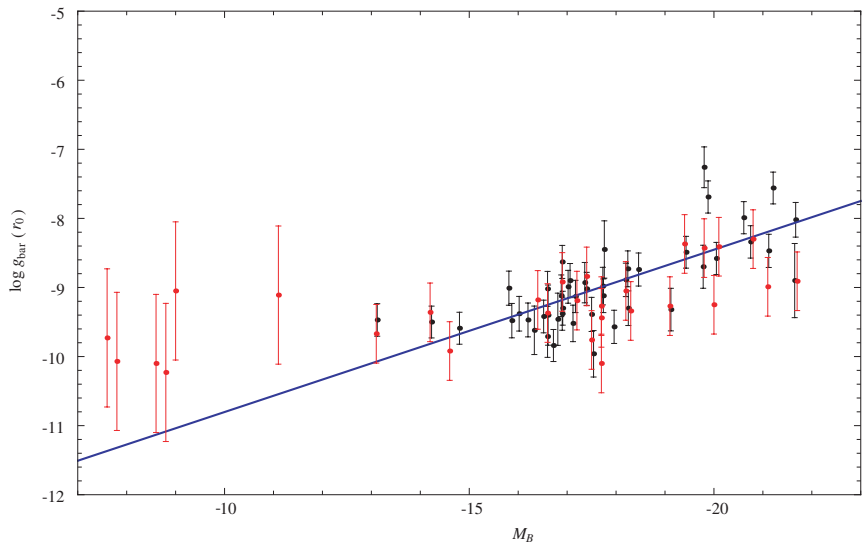}
\caption{Our best fit $g_{bar}(r_0)$ vs $M_B$ relation superimposed to the G09 (red) and our (black) acceleration data.}
\label{fig: gbarfigs}
\end{figure*}


In Fig. 3c, we plot the surface density of dark matter for a system composed only of dark matter, as in dissipationless simulations.
In order to get the surface density just of dark matter, we did not take into account baryons in our SIM. 
In this case, we get $\rho_0 r_0 \propto M^{0.20 \pm 0.05}$.

%
%

As reported in the introduction, also NRT10 questions D09 and G09 results, who claimed a universal DM projected density for all galaxy types and masses (see also Kormendy \& Freeman 2004). They reproduce their results reasonably well, finding a fairly constant surface density on average across a large mass range of dwarfs and late types, but also suggesting a systematic increase with mass as found by B09.
One objection that could arise at this point is that the results presented in Fig. 3, are not better fits to D09, and G09 data than a costant value.  
Our result is not a fit to the data, but it is the result of a theoretical model, which predicts, for example in the case of the surface density of baryons $\alpha=0.18 \pm 0.05$, and equal error bars in the other cases. Our result with an error of $\pm 0.05$ and not $\pm 0.2$ (as in D09, and G09), differs at a level of 3 $\sigma$ from G09 (and D09) result, and specify among all the slopes permitted by G09 ($[-0.2,0.2]$) a smaller set. This has a strong physical consequence, namely that the surface density is not a universal quantity.      
In fact, in order the acceleration scales $g_i(r_0)$ (or the surface density), (with $i = \{DM, bar\}$) was an universal quantity as argued by D09 and G09, one should find no correlation with any stellar or DM property. The wide scatter in the slope obtained by the authors ($0 \pm 0.2$) prejudice this conclusion, since it implies at least a correlation with $M_B$. D09, and G09 results are compatible with a flat slope implying no correlation, but are also compatible with a non flat slope implying correlation with $M_B$. Our result is not compatible with a flat slope. 

{
Nevertheless the previous discussion, one could argue that even if the model is consistent, it is not necessarily more correct than the constant.  }

In order to further clarify the previous issue, we repeated G09, and D09 calculations in a similar way they did, using a Burkert's profile fit, but for a larger sample. 
{The details of the analysis: the data used, the way the fits were performed, were published in Cardone \& Del Popolo (2012). 
Here, we summarize the main features of the paper. In the quoted paper, we searched the literature for systems with high quality rotation curve data probing the gravitational potential with good sampling, large radial extension (i.e., up to $R > R_{opt} = 3.2 R_d$ with $R_d$ the disk scale length) and small errors. 
In order to reduce the potential impact of beam smearing on the circular velocity, we used, whenever possible, H$\alpha$ data or a combination of HI and H$\alpha$.
Our sample contains 58 galaxies (also containing G09 and D09 galaxies) from different sources (de Blok \& Bosma 2002; Simon et al. 2005; Spano et al. 2008; THINGS; Oh et al. (2008); Swaters et al. 2011)
spanning nine orders of magnitude in the $B$ band luminosity (roughly, $-22 \le {\cal{M}}_B \le -13$) and different spiral classes (from dwarfs system to low and high surface brightness galaxies). 
In order to be consistent with G09, we assumed a Burkert model to fit the rotation curve.

The model parameters were obtained through a likelihood analysis by maximizing ${\cal{L}}({\bf p}) \propto \exp{[-\chi^2({\bf p})/2]}$ with

\begin{equation}
\chi^2 = \sum_{i = 1}^{{\cal{N}}}{\left [ \frac{v_c^{obs}(r_i) - v_{c}^{th}(r_i, {\bf p})}{\sigma_i} \right ]^2}
\label{eq: defchisq}
\end{equation}
where $v_c^{obs}(r_i)$ is the measured circular velocity for the $i$\,-\,th point (with $\sigma_i$ the error), $v_c^{th}(r_i, {\bf p})$ the theoretically predicted value for the given set of parameters ${\bf p} = (\Upsilon_{\star}, V_{vir}, c_{vir})$ and the sum is over the ${\cal{N}}$ data points.  
A Markov Chain Monte Carlo (MCMC) method running three chains checking the convergence through the Gelman \& Rubin (1992) criterion was used to efficiently explore the quoted 3D space of parameters.

After having constrained the halo model parameters and the disc stellar $M/L$ ratio, the Newtonian acceleration of both the baryon (disc\,+\,gas) and DM components can be estimated through Eqs. (1) and (4) in Cardone \& Del Popolo (2012). The uncertainties on the fitting parameters and the correlations among them was obtained evaluating $g_i(r_j)$ along the Markov chain for each galaxy and use the resulting values to estimate the median and $68\%$ confidence ranges. Note that, here, the label $i$ denotes the galaxy component (baryons or DM), while $j$ refers to the reference radius adopted. 

Our method has two differences when compared to G09. Firstly, we fit the full rotation curve with the Burkert model thus being also able to check whether or not the model well reproduces the observed data. 
G09 did actually not perform any fit to the rotation curve data, but used literature data and an approximate law to convert the scale length of the model used by other authors to their Burkert model $r_0$ radius. This choice does not automatically ensure that the theoretically predicted rotation curves well match also in the outer regions. 
However, the scaling law used by G09 indeed provides values of $g_{i}(r_0)$ consistent with our ones with $\langle \Delta \log{g_{DM}(r_0)} \rangle = -0.03$ and  $\langle \Delta \log{g_{bar}(r_0)} \rangle = -0.02$ as estimated by the 18 galaxies common to both samples\footnote{We miss 10 galaxies from the G09 sample because of different reasons. }

A second difference with G09 concerns the estimate of the uncertainties. As previously reported, we have determined (rather than assumed) the halo model parameters and the stellar $M/L$ ratio, and this allowed us to attach to each $g_i(r_0)$ value an error fully taking into account both the fitting parameters uncertainties and their correlations. 
We then add in quadrature a systematic error related to the uncertainties on the method used to determine $g_{i}(r_0)$. This was obtained  
comparing with the G09 values for the galaxies common to both samples and add in quadrature $0.23$ to the statistical errors on $\log{g_{i}(r_0)}$ (see Cardone \& Del Popolo (2012) for details).


In order to check the universality hypothesis, we fit the following log\,-\,linear relations

\begin{displaymath}
\log{g_i(r_0)} = a \log{\frac{M_{vir}}{10^{11} \ {\rm M_{\odot}}}} + b \ ,
\end{displaymath}

\begin{displaymath}
\log{g_i(r_0)} = a M_B + b \ ,
\end{displaymath}
using the Bayesian method, getting (we show here just the $g_{bar}-M_B$ case) 
for $g_{bar}-M_B$, $a=0.231^{+0.029}_{-0.027}$, $b=13.04^{+0.064}_{-0.038}$, and the intrinsic scatter $\sigma_{int}=0.23^{+0.06}_{-0.06}$ obtained through the D'Agostini (2005) Bayesian method. 

In Fig. 4, we plot our data (black) superimposed to G09 (red) and the best fit line.

%
%

Fig. 4 shows that in the range $-20 \le M_B \le -15$, our data are well superimposed to the G09 ones except for few cases where our estimates is larger than that in G09\footnote{Since the values in G09 comes from scaling the result for other halo models to the Burkert one, while our own are directly based on fitting this halo profile to the rotation curve data, we consider our values more reliable. }, and that if we exclude the galaxies with $\log{g_{bar}(r_0)} > -8.0$ and add the Milky Way dwarfs with $M_B > -12$, a constant $g_{bar}(r_0)$ would be preferred. However, the situations changes when taking into account the four systems with $\log{g_{bar}(r_0)} > -8.0$ (namely, NGC\,2841, NGC\,3521, NGC\,4736, NGC\,6946), which belong to the THINGS sample so that they have a high quality rotation curve and a not small inclination (as estimated from the HI data). These systems were not considered in G09.  
The reason for the discrepancy on the constancy of $g_{bar}(r_0)$ with $M_B$ is mainly due to our sample including galaxies with larger acceleration values. 
The previous result clearly shows the non-universality of the quoted relation.


}

\section{Conclusions}

In the framework of the SIM model of DP09, in this paper the correlation between the central surface density and the halo core radius of galaxy, and cluster of galaxies, dark matter haloes was analyzed. The main result is that $r_\ast$ (the halo characteristic radius)is not an universal quantity as claimed by D09 and G09.  
We obtained a surface density , $S \propto M^{0.16 \pm 0.05}$, and $S \propto M^{0.18 \pm 0.05}$, for masses $> 5 \times 10^{10} M_{\odot}$, for the cases examined in D09 and G09, respectively. Our results suggest that the constant surface density, claimed by the previously quoted authors, is probably unlikely. 


\section*{Acknowledgments}

ADP is partially supported by a visiting research fellowship from FAPESP (grant 2011/20688-1), and wishes also to thank the Astronomy Department of Sao Paulo University for the facilities and ˜
hospitality.

\end{document}